 \pdfoutput=1
   \documentclass[fleqn,usenatbib,usedcolumn]{mnras}
   \usepackage[british]{babel}             
   \usepackage{txfonts}                    
   \let\la=\lesssim 
   \usepackage[T1]{fontenc}                
   \usepackage{graphicx}                   
\title[Short- and mid-term variations in the frequencies]{A thorough analysis of the short- and mid-term activity-related variations in the solar acoustic frequencies}

\author[A. R. G. Santos et al.]{A. R. G. Santos$^{1,2,3}$\thanks{E-mail: asantos@astro.up.pt; mcunha@astro.up.pt},
M. S. Cunha$^{1,2}${\color{blue}\footnotemark[1]},
P. P. Avelino$^{1,2}$,
W. J. Chaplin$^{3,4}$,
and T. L. Campante$^{3,4}$
\\
$^{1}$ Instituto de Astrof\'{i}sica e Ci\^{e}ncias do Espa\c{c}o, Universidade do Porto, CAUP, Rua das Estrelas, PT4150-762 Porto, Portugal\\
$^{2}$ Departamento de F\'{i}sica e Astronomia, Faculdade de Ci\^{e}ncias, Universidade do Porto, Rua do Campo Alegre 687, PT4169-007 Porto, Portugal\\
$^{3}$ School of Physics and Astronomy, University of Birmingham, Edgbaston, Birmingham B15 2TT, UK\\
$^{4}$ Stellar Astrophysics Centre (SAC), Department of Physics and Astronomy, Aarhus University, Ny Munkegade 120, 8000 Aarhus C, Denmark
}

\pubyear{2016}

\begin{document}
\label{firstpage}
\pagerange{\pageref{firstpage}--\pageref{lastpage}}
\maketitle

\begin{abstract}
{The frequencies of the solar acoustic oscillations vary over the activity cycle. The variations in other activity proxies are found to be well correlated with the variations in the acoustic frequencies. However, each proxy has a slightly different time behaviour. Our goal is to characterize the differences between the time behaviour of the frequency shifts and of two other activity  proxies, namely, the area covered by sunspots and the 10.7cm flux. We define a new observable that is particularly sensitive to the short-term frequency variations. We then compare the observable when computed from model frequency shifts and from observed frequency shifts obtained with the Global Oscillation Network Group (GONG) for cycle 23. Our analysis shows that on the shortest time-scales the variations in the frequency shifts seen in the GONG observations are strongly correlated with the variations in the area covered by sunspots. However, a significant loss of correlation is still found. We verify that the times when the frequency shifts and the sunspot area do not vary in a similar way tend to coincide with the times of the maxima of the quasi-biennial variations seen in the solar seismic data. A similar analysis of the relation between the 10.7cm flux and the frequency shifts reveals that the short-time variations in the frequency shifts follow even more closely those of the 10.7cm flux than those of the sunspot area. However, a loss of correlation between frequency shifts and 10.7cm flux variations is still found around the same times.}
\end{abstract}




\section{Introduction}\label{sec:intro}

The frequencies of the solar acoustic modes vary periodically and in phase with the magnetic activity over the so-called 11-yr solar cycle \citep[e.g.][]{Woodard1985,Libbrecht1990a,Elsworth1990,Chaplin1998,Dziembowski2005,Tripathy2011,Salabert2011,Salabert2015,Howe2015}. Activity-related frequency shifts were also detected in three solar-like stars \citep[][]{Garcia2010,Salabert2016,Regulo2016} and are expected to be common in solar-type pulsators.

It has been argued that the frequency shifts result from variations of both the weak and strong components of the solar magnetic field \citep[][]{Tripathy2007,Jain2009,Broomhall2015,Santos2016}. This is corroborated by their strong correlation with the 10.7cm flux \citep{Chaplin2007,Tripathy2007,Jain2009,Broomhall2012,Simoniello2012}, which is mostly sensitive to the weak component, but also affected by the strong component of the magnetic field \citep[e.g.][]{Covington1969,Tapping1987,Tapping1990}, as well as by their sensitivity to the latitudinal distribution of active regions \citep[e.g.][]{Hindman2000,Howe2002,Chaplin2004,Broomhall2012} and their strong correlation with the sunspot areas \citep[][]{Jain2012,Broomhall2015}. Still, different studies have shown a decrease in the correlation between the activity proxies and the frequency shifts around the maximum of the solar cycle \citep[e.g.][]{Jain2009,Simoniello2012,Broomhall2015}.

In addition to the 11-yr variation, the solar acoustic frequencies also show a quasi-biennial variation \citep[][]{Fletcher2010,Broomhall2012,Simoniello2012,Simoniello2013,Broomhall2015}, which is strongly correlated with the mid-term periodicities found in other activity proxies \citep[][]{Broomhall2012,Broomhall2015}. The signature of this quasi-biennial variation in the frequencies is seen over all phases of solar activity. However, its amplitude is highest around the solar maximum, suggesting that it is modulated by the 11-yr cycle. These mid-term frequency shifts show a weaker dependence on the frequency than the longer term (11-yr) frequency shifts \citep[][]{Fletcher2010,Broomhall2012,Simoniello2013}, which has been interpreted as an indication that they originate from changes in layers that are located deeper than those inducing the 11-yr signal.

One of the difficulties in interpreting the correlations found between variations in the different activity proxies and variations in the oscillation frequencies comes from the fact that the latter result from the combination of different physical phenomena. In fact, the oscillation frequencies are sensitive to the direct effect of the magnetic field, which might be significant in active regions, but also to variations in the solar structure and dynamics that are induced both by the weak and strong magnetic field components. Luckily, these phenomena do not  all have the same characteristic time-scale, and thus, by considering the behaviour of the frequency shifts  on short-, mid-, and long-terms one may hope to move forward in the interpretation of the observed correlations. 

In this work, we propose a new observable to investigate the short- and mid-term variations of the frequency shifts. We argue that this  observable, based on a weighted sum of frequency-shift variations, is insensitive to the long-term cycle variations, allowing us to amplify the signature of the short-term variations contained in the data. 

\section{Observed and model frequency shifts}\label{sec:data}

In order to study the short-term variations of the solar acoustic frequencies one must consider observed frequency shifts obtained from relatively short time series, such as those derived by \citet{Tripathy2011} using data from the Global Oscillation Network Group (GONG). In that work, the authors computed oscillation frequencies and the corresponding frequency shifts with a cadence of 36 d and an overlap of 18 d. These observations, thus, provide two datasets that we will analyse separately, each composed of a sequence of consecutive, independent, frequency shifts obtained from 36-d time series: Sample 1 and Sample 2 -- starting, respectively, on the first and second data point of the original dataset from \citet{Tripathy2011}.

\begin{figure}\centering
\includegraphics[trim=12 0 0 0mm,clip,width=\hsize]{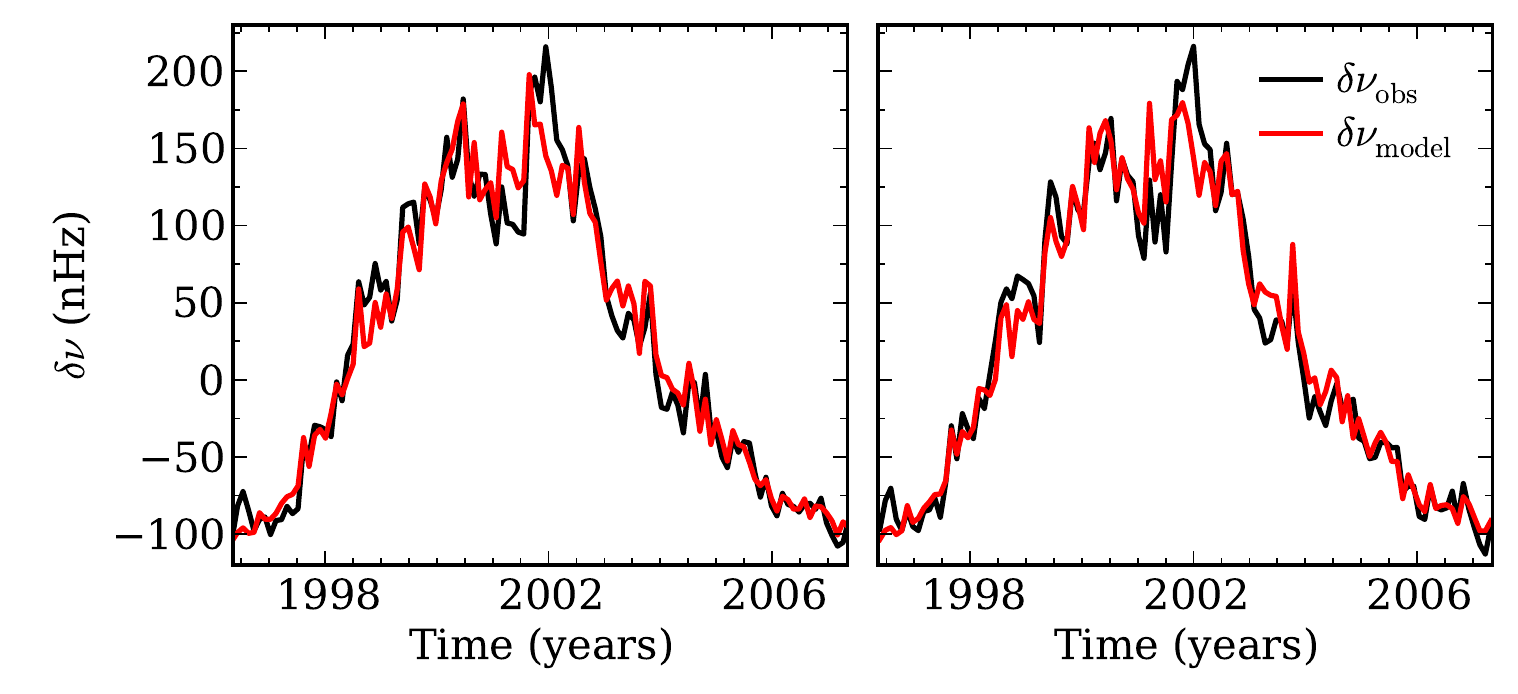}
\caption{Comparison between the observed (black) and model (red) frequency shifts over the solar cycle 23 obtained using Samples 1 and 2 (left and right panel, respectively -- see text for details).}\label{fig:fshifts}
\end{figure}

For, the model frequency shifts we use a parametrized model proposed by \citet[hereafter Paper I]{Santos2016}. Accordingly, the model frequency shifts are described by the sum of two components
\begin{equation}
\delta\nu_{\rm model}=\delta\nu_{\rm spots}+\delta\nu_{\rm global},\label{eq:model}
\end{equation}
where the first term on the r.h.s of equation (\ref{eq:model}) represents the frequency shifts induced by sunspots (essentially proportional to the variation of the total area covered by sunspots) and the second represents a smooth 11-year variation induced by other phenomena, such as the variation of the global magnetic field and structural and thermal changes acting on a solar-cycle time-scale. The spot-induced component, $\delta\nu_{\rm spots}$, is derived from the daily sunspot records of the National Geophysical Data Center, part of the National Oceanic and Atmospheric Administration (NGDC/NOAA; www.ngdc.noaa.gov), while the long-term smooth component is obtained by fitting the observed frequency shifts with the function proposed by \citet{Hathaway1994} to describe the sunspot number over the solar cycle. By comparing the total model frequency shifts with the frequency shifts observed over the solar cycle 23 \citep[from][]{Tripathy2011}, \citet{Santos2016} found that the shortest-term variations seen in the GONG observational data are well described by the spot-induced frequency shifts, which they estimate to account for about 30 per cent of the long-term variations observed throughout the solar cycle. The comparison between the model predictions and the GONG data for the two observational sets of independent frequency shifts is shown in Fig. \ref{fig:fshifts}\footnote{A similar comparison was shown in Paper I, but for the  two datasets combined.}.

A quasi-biennial variation with an 11-year amplitude modulation was also identified by the authors in the residuals of the model-data comparison. This component, as well as the 11-year variation, has been identified previously in data obtained from different space and ground-based instruments and discussed in a number of works (cf. Section \ref{sec:intro}), while the shortest-term component (varying on a time-scale of days) was first discussed in Paper I. 

\section{Description of the method}
\label{sec:cortest}

\begin{figure}
\includegraphics[width=\hsize]{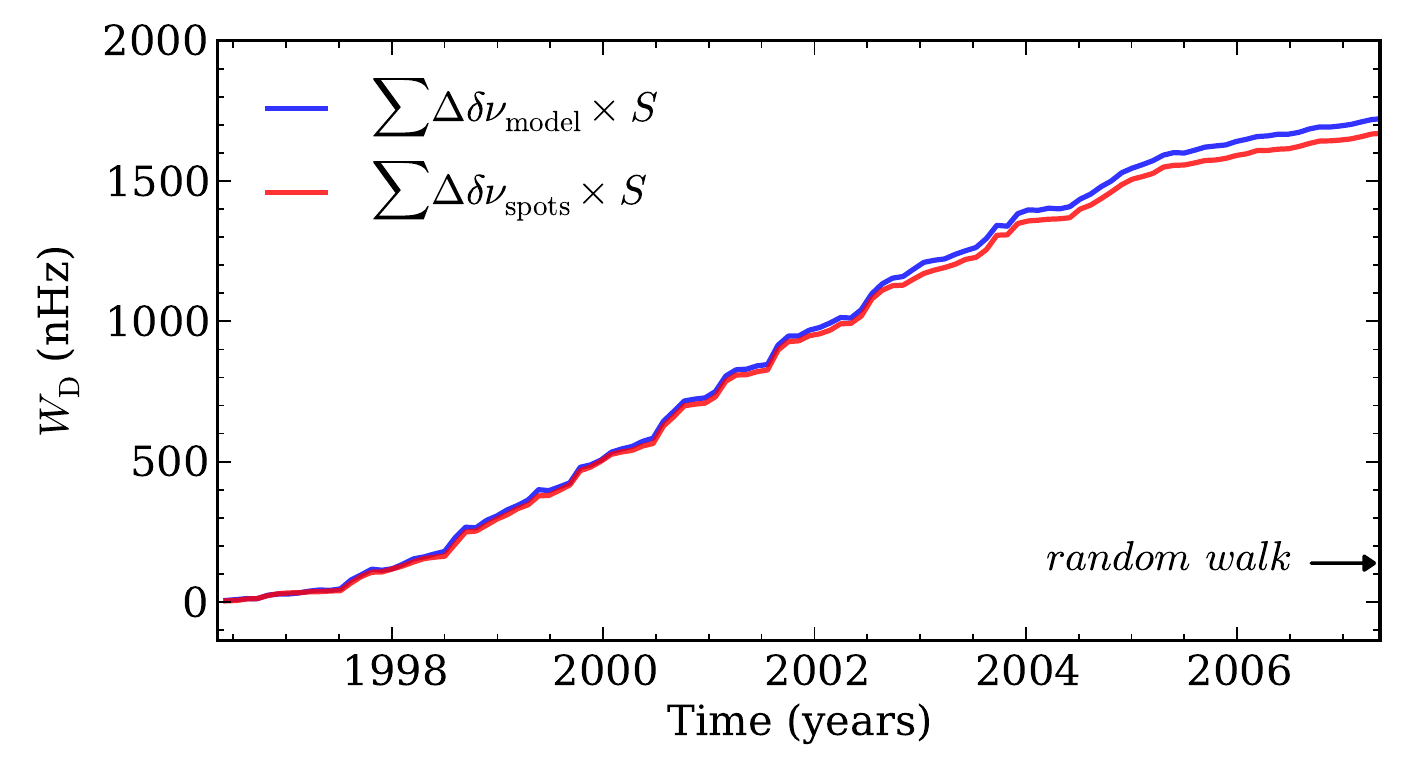}
\caption{Weighted sum of the variations in the model (blue) and spot-induced (red) frequency shifts (derived from the observational sunspot data), where the weight $S$, in equation (\ref{eq:cortest}), is determined by the variation in the area covered by sunspots. The black arrow marks the expected standard deviation for a random walk.}\label{fig:modelcortest}
\end{figure}

The variations in the solar acoustic frequencies are known to be correlated with the variations in the solar activity indicators, in particular with the area covered by sunspots. The latter varies on a time-scale of days and induces frequency variations on a similar time-scale. Unfortunately, it is not possible to derive accurate frequency shifts from time-series as short as a few days. Nevertheless, such short time-scale variations are expected to be the main source of the 36-d frequency variations seen in the GONG data. To test this possibility and investigate more closely the short-term frequency shifts, we define a new observable 
\begin{equation}
W_{\rm D}=\sum_k\Delta\delta\nu_k\times S_k,\label{eq:cortest}
\end{equation}
as the weighted sum of the frequency shift differences. Here $\Delta\delta\nu_k$ is the difference between the frequency shifts measured in two consecutive data bins (associated to times $k$ and $k-1$, i.e. $\Delta\delta\nu_k=\delta\nu_k-\delta\nu_{k-1}$), thus corresponding to the bin-to-bin frequency variation. Also, $S$ is a weight to be determined according to the variation of the area covered by sunspots. At a given time $k$, if the total area covered by spots increases with respect to its value at the time $k-1$, the weight, $S_k$, will be $1$, otherwise $S_k=-1$. If the short-term frequency-shift differences are indeed strongly correlated with the variations in the area covered by spots, as predicted by the model for the spot-induced frequency shifts, they will be summed positively when performing this weighted sum. Thus, the new variable $W_{\rm D}$ is expected to be most sensitive to the short-term variation component of the frequency shifts, with the smooth, long-term component, being essentially cancelled due to the varying positive and negative weights. To show that this is indeed the case, we compute $W_{\rm D}$ using the spot-induced frequency shifts alone ($\delta\nu_{\rm spots}$), derived from the observational sunspot data, and using the total model frequency shifts ($\delta\nu_{\rm model}$,which includes the spot contribution and the long-term smooth component, cf. equation (\ref{eq:model})), where the weight $S$ is determined from the observed sunspot areas (NGDC/NOAA). The comparison between the two cases is shown in Fig. \ref{fig:modelcortest}, where the maximum difference between the two curves is one order of magnitude smaller than what would be found if the large-scale variations were summed positively, thus confirming that the smooth, long-term component almost cancels out when computing $W_{\rm D}$ for the 36-d cadence. The arrow marks the expected standard deviation at the end of the solar cycle for the case in which the frequency shift variations are described by a random walk, thus providing an indication of the interval where the maximum value of the curves would be expected to lie, if the frequency-shift differences were completely uncorrelated with the variations in the area covered by sunspots. The fact that the maxima of the two curves is so much greater than that value is a consequence of the almost perfect correlation between the spot-induced frequency-shift differences predicted by the model and the corresponding variations in the area covered by spots.

\begin{figure*}
\centering
\includegraphics[trim= 65 0 0 0 mm, clip,width=0.9\hsize]{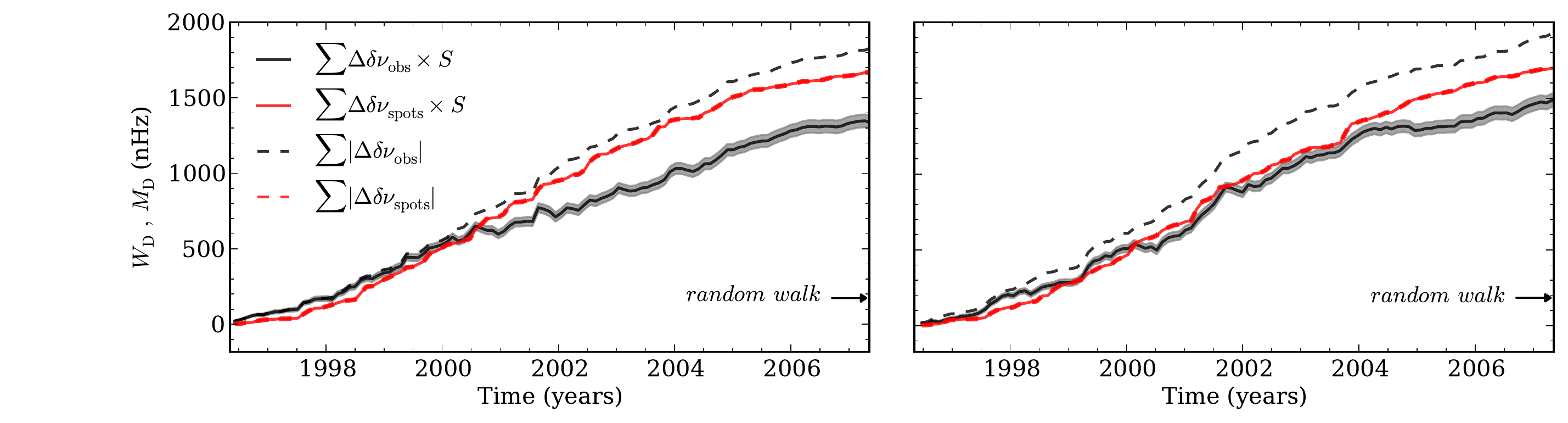}
\caption{Left and right panels correspond to the results obtained for Samples 1 and 2, respectively. Weighted sum $W_{\rm D}$ of the variation in the observed (black solid line) and spot-induced (red solid line) frequency shifts (derived from the observational sunspot data), where the weight, $S$, is determined by the variation in the total observed area covered by sunspots. The grey region represents the $1\sigma$ confidence interval from the black continuous curve resulting from the errors in the observed frequency shifts. The dashed lines represent the sum, $M_{\rm D}$, of the absolute (modulus) values of the frequency-shift differences. The arrow marks the expected standard deviation for a random walk.}\label{fig:resobs}
\end{figure*}

We now compute $W_{\rm D}$ for the Samples 1 and 2 of the observed frequency shifts obtained from GONG data and compare the results with the hypothetical case of complete correlation, defined as the case in which the variations in the area covered by sunspots and in the frequency shifts always have the same sign (meaning that an increase (decrease) in one always correspond to an increase (decrease) in the other). In that case the weighted sum defined by equation (\ref{eq:cortest}) reduces to the sum of the absolute (modulus) values of the frequency-shift differences ($M_{\rm D}$)
\begin{equation}
M_{\rm D}=\sum_k|\Delta\delta\nu_k|.
\end{equation}
The quantities $W_{\rm D}$ (black solid line) and $M_{\rm D}$ (black dashed line) obtained from Samples 1 and 2 (left and right panels, respectively) are shown in Fig. \ref{fig:resobs}. The comparison between the maximum of the black solid line and the expected standard deviation for a random walk allows us to conclude that a significant correlation exists between the sunspot-area variations and the frequency-shift differences. In fact, in both cases the black continuous curve reaches a maximum value that is $7\sigma$ above that of a random walk, meaning that if the short-term variations in the frequency shifts and in the areas were uncorrelated, the probability of obtaining the deviation reached by the black solid lines would be less than $10^{-11}$. For comparison we show in the same plot the 
the results for the model spot-induced frequency shifts derived from the observational sunspot data. The red solid and dashed lines correspond to the quantities $W_{\rm D}$ and $M_{\rm D}$, respectively, obtained from Samples 1 and 2 (left and right panels, respectively). In this case, the correlation is perfect and, unlike in the case of the observed frequency shifts, the solid and dashed curves overlap.

We also verified to what extent the results for $W_{\rm D}$ are affected by the error associated to the observed frequency shifts. By taking the observational data points and assuming that the corresponding errors are Gaussian, we obtain random samples for the frequency shifts. We then estimate the standard deviation from $W_{\rm D}$, which is represented by the grey area in Fig. \ref{fig:resobs}. For some data points, the observational errors are of the same order as the frequency-shift differences. However, those data points do not contribute significantly to $W_{\rm D}$. Clearly, the uncertainty in the black curve (grey region) associated to the observational errors in the frequency shifts does not explain the loss of correlation found.

We thus conclude that a strong, but not complete correlation exists between the short-term variations in the frequency shifts and in the spots' areas. The origin of the observed loss of correlation can be twofold: it may result from our inability to compute the true correlation between these two variations, or it may correspond to a genuine loss of correlation introduced either by changes in the effect that sunspots have on the frequencies or by the effect of other physical phenomena (besides spots) which may influence the solar frequency shifts on short time-scales. In the next section we will discuss a potential source for that loss of correlation, associated to the lack of information about the sunspots on the invisible side of the Sun.

\section{Impact of ignoring the solar invisible side on the computation of $W_{\rm D}$}\label{sec:simul}

Even if a complete correlation existed between the variations in the frequency shifts and in the total area covered by spots, we would still expect to see a significant loss of correlation when comparing the quantities $W_{\rm D}$ and $M_{\rm D}$ derived from the observations. The reason is that the sunspot areas we consider for the computation of $W_{\rm D}$, obtained from the NGDC/NOAA databases, concern only the visible sunspot groups, while the acoustic frequencies are affected by sunspot groups emerging over the whole solar surface. In this section, we estimate the effect of having ignored the invisible sunspot groups in the computation of $W_{\rm D}$ of the previous section. To that end, we use synthetic sunspot data obtained with an empirical tool developed by \citet{Santos2015}. This tool generates synthetic sunspot daily records analogous to the real sunspot data from NGDC/NOAA, starting from simulations of sunspot groups on both the visible and invisible sides of the Sun. 

Using the model presented in Paper I, we compute the spot-induced frequency shifts for the synthetic sunspot data, $\delta\nu_{\rm synthetic}$, taking into account all the spots, including those appearing on the visible and invisible sides of the Sun. Since the model for the frequency shifts is parametrized, it needs to be calibrated by comparison with the observed data. In this case, we are interested in analysing the short-term variations in the frequency shifts. We thus have calibrated the frequency shifts such as to make their absolute sum equal to the absolute sum of the observed counterparts, i.e. $\sum_k|\Delta\delta\nu_{\rm obs}|_k=\sum_k|\Delta\delta\nu_{\rm synthetic}|_k$.
We then compute $W_{\rm D}$ for the synthetic data, by combining the variations in the frequency shifts obtained from all the synthetic sunspot groups with the weights $S_k$ computed only from the visible synthetic sunspot areas. Since the sunspot cycle simulations are stochastic, we repeated this exercise 1000 times, computing the values of $W_{\rm D}$ and $M_{\rm D}$ for all simulations.  

\begin{figure}
\centering
\includegraphics[width=\hsize]{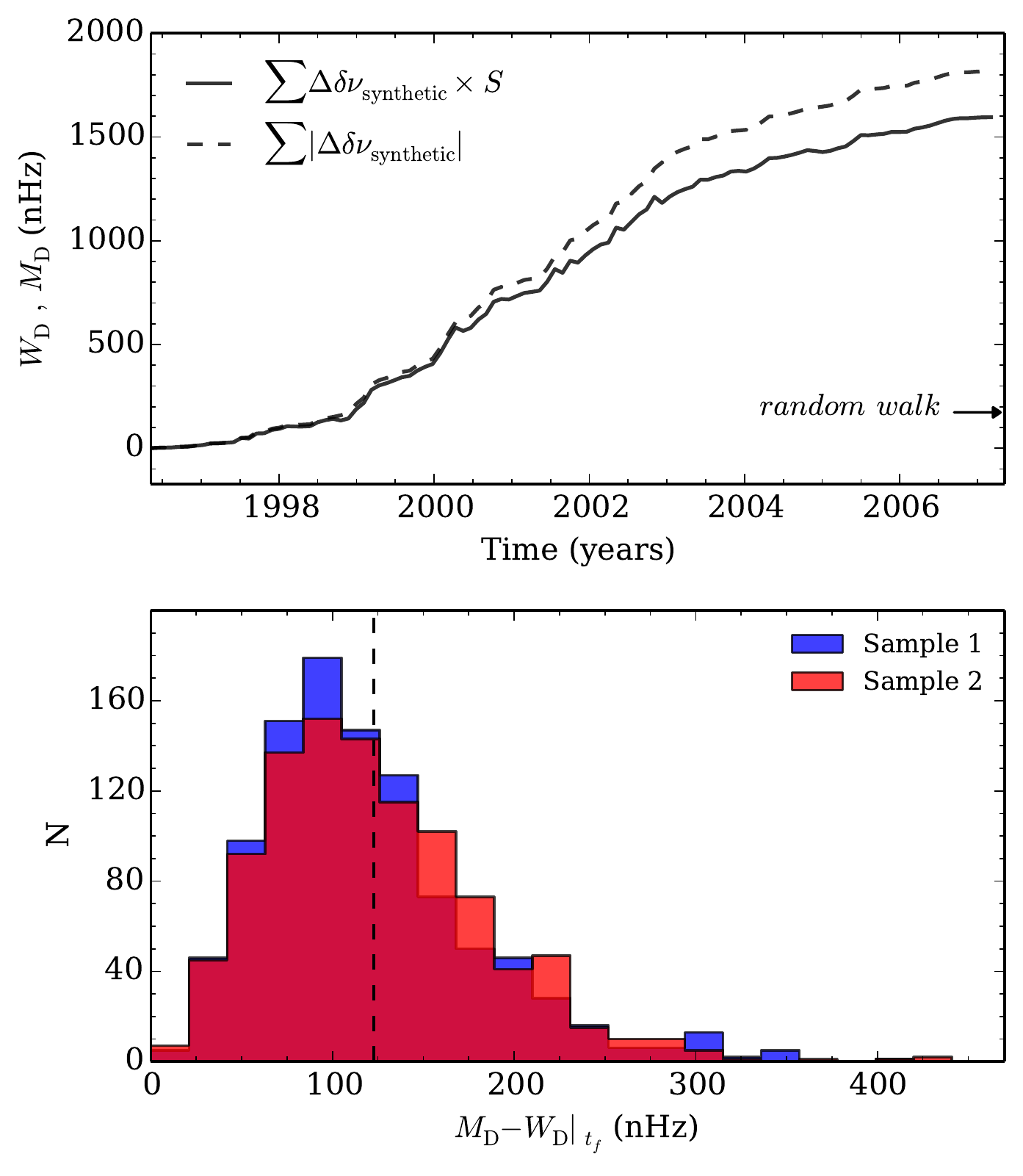}
\caption{Top panel: Comparison between the quantities $M_{\rm D}$ (dashed line) and $W_{\rm D}$ (solid line) for a particular synthetic cycle. The arrow marks the expected standard deviation for a random walk. Bottom panel: Distribution of the difference between the quantities $M_{\rm D}$ and $W_{\rm D}$ at the end of the synthetic solar cycle (at $t_f$). The dashed line marks the averaged difference found for both samples of independent frequency shifts.}\label{fig:distsimul}
\end{figure}

The quantities $M_{\rm D}$ and $W_{\rm D}$ derived from the synthetic data are shown in the top panel of Fig. \ref{fig:distsimul} for one particular simulation. The difference between these quantities gives an estimate of the loss of correlation that can be explained by our not accounting for the sunspot groups on the invisible side of the Sun when computing the weights $S_k$. The bottom panel of Fig. \ref{fig:distsimul} shows the distribution of that difference at the end of the solar cycle. Taking the average of the distribution as an indicator and comparing it with the difference between $M_{\rm D}$ and $W_{\rm D}$ computed in Section \ref{sec:cortest} for the observations, we conclude that only about 30 per cent of the loss of correlation found in the analysis of the real data may be explained by this non-physical effect. Therefore, a significant part of the loss of correlation detected in Section \ref{sec:cortest} is expected to have its origin in physical effects other than sunspots that affect the short-term frequency sifts and that are not, themselves, fully correlated with the total area covered by sunspots. 

\section{Epochs of distinct behaviour between the frequency shifts and the sunspot areas}\label{sec:noncor}

\begin{figure*}\centering
\includegraphics[width=0.87\hsize]{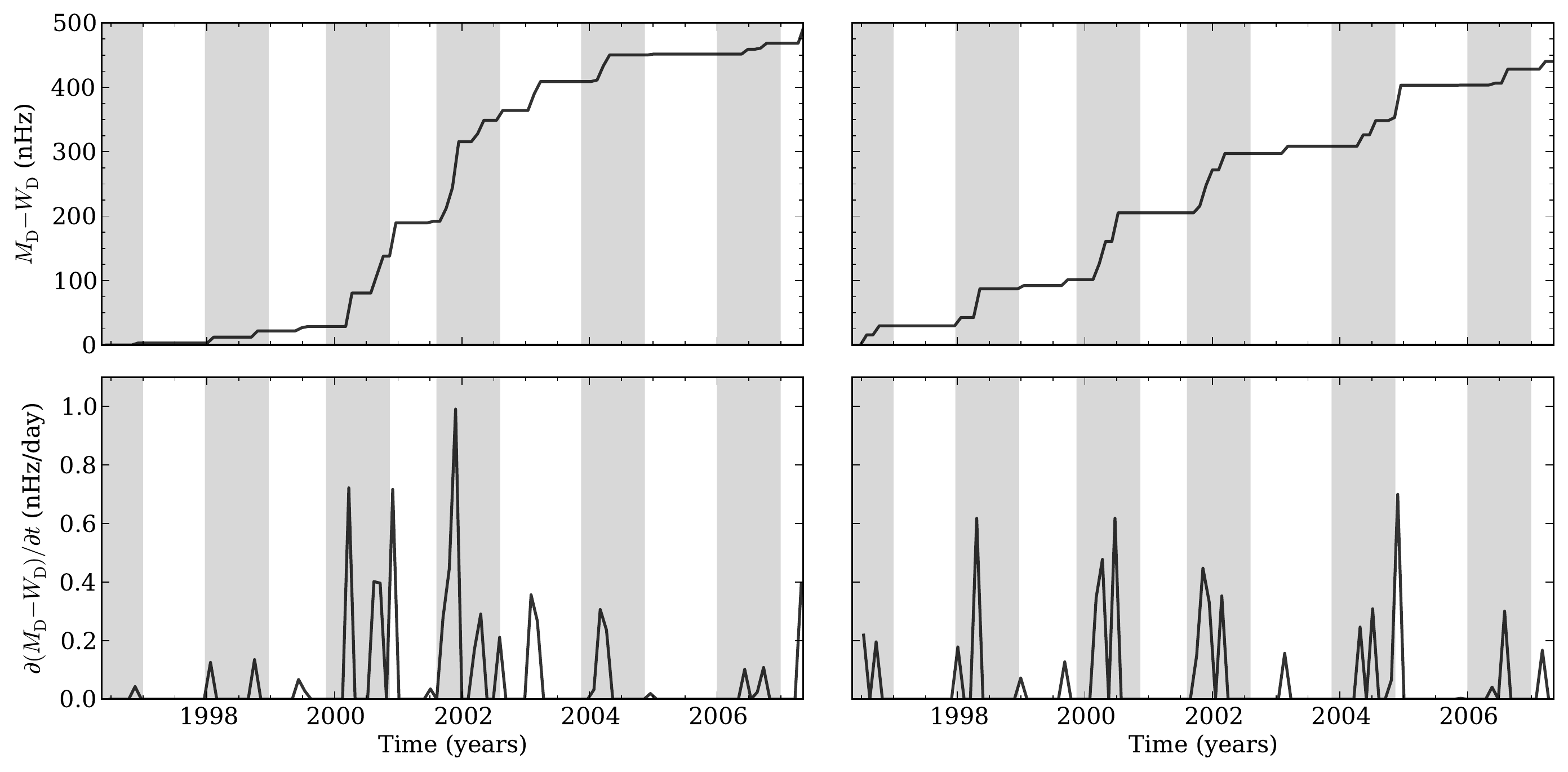}
\caption{The left panel concerns Sample 1, while the right panel corresponds to Sample 2 (see text for details). Top panels: Difference between the absolute sum, $M_{\rm D}$, and the weighted sum, $W_{\rm D}$, of the observed frequency-shifts variations, i.e. between the black-dashed and solid lines in Fig. \ref{fig:resobs}. Bottom panels: Time derivative of the difference shown in the top panels. The grey bars, with a width of one year, are centered around the approximate location of the maxima of the quasi-biennial signal detected by \citet{Broomhall2012}.}\label{fig:diffderiv}
\end{figure*}

While the frequency-shift differences are found to be strongly correlated with the variations in the area covered by sunspots, there are particular epochs when the frequency shifts and the sunspot areas vary in the opposite way, leading to the loss of correlation found in Section \ref{sec:cortest}. In this section, we identify those epochs.

The top panels of Fig. \ref{fig:diffderiv} show the difference between $M_{\rm D}$ (black dashed line in Fig. \ref{fig:resobs}) and $W_{\rm D}$ (black solid line in Fig. \ref{fig:resobs}) for the observed frequency shifts. The results for the two samples differ slightly. For sample 1 (left panel) the differences have a larger amplitude around the solar maximum, while for sample 2 they are more evenly spread. We recall that the two samples provide 36-d averages of the frequency shifts that are shifted by 18-d. The differences seen in the results for the two samples highlight that a significant loss of correlation takes place on shorter times scales than 18 d. In the discussion that follows we will, therefore, consider only aspects that are identified in both samples.

The bottom panels of Fig. \ref{fig:diffderiv} show the derivative of the difference between $M_{\rm D}$ and $W_{\rm D}$ with respect to the time. This quantity is zero when the variations in the frequency shifts have the same sign as those in the sunspot areas, differing from zero when the two variations have an opposite sign. The most significant correlation losses occur around the same times for the two samples of independent frequency shifts. The peaks are quasi-periodic, coinciding with the maxima of the quasi-biennial periodicity observed in the solar acoustic frequencies. This is illustrated by the grey bars in Fig. \ref{fig:diffderiv}, which have a width of 1 year and are centered at the locations of maxima of the quasi-biennial signal found by \citet{Broomhall2012}.

\section{Discussion}\label{sec:discussion}

Our results show that there is a strong correlation between the short-term variations in the frequency shifts and in the area covered by sunspots. However, the loss of correlation between the two is still significant and cannot be fully explained by the combined effect of the sunspot groups on the invisible side of the Sun and of the observational error in the observed frequency-shift differences. Moreover, the detailed analysis of the difference between the quantities $W_{\rm D}$ and $M_{\rm D}$ shows that opposite variations in the frequency shifts and in the sunspot area are more significant around the times of maxima of the quasi-biennial signal. This clear signature of the quasi-biennial variations in the quantity $W_{\rm D}$ highlights the fact that the short time-scale variations are modulated on a quasi-biennial time-scale.

The above findings could point to a quasi-biennial change in the effect of the sunspots. However, we find this possibility difficult to understand on physical grounds because it would require that an increase in the sunspot area led to a decrease in the frequency shifts at the maximum of the quasi-biennial variations. According to the model for the spot-induced frequency shifts (Paper I), this would imply that the phase shift induced by a sunspot on the acoustic wave travelling through it changed sign from the minimum to the maximum of the quasi-biennial variation. Another, perhaps more likely possibility is that other active-related features, such as plage, not accounted for in the computation of $W_{\rm D}$, could  contribute to  the short-term variations in the frequency shifts. However, given the asymmetry found between the behaviour of the correlations at the maximum and minimum of the quasi-biennial variations, their effect would have to dominate the short-term frequency-shifts variations during the former, but be unimportant, compared to the effect of sunspots, during the latter. This possibility will be further investigated by inspecting the behavior 10.7 cm flux, discussed in the next subsection.

\subsection{Short-term variations in the 10.7cm flux}\label{sec:F10.7}

\begin{figure*}\centering
\includegraphics[trim=50 50 0 0mm,clip,width=0.9\hsize]{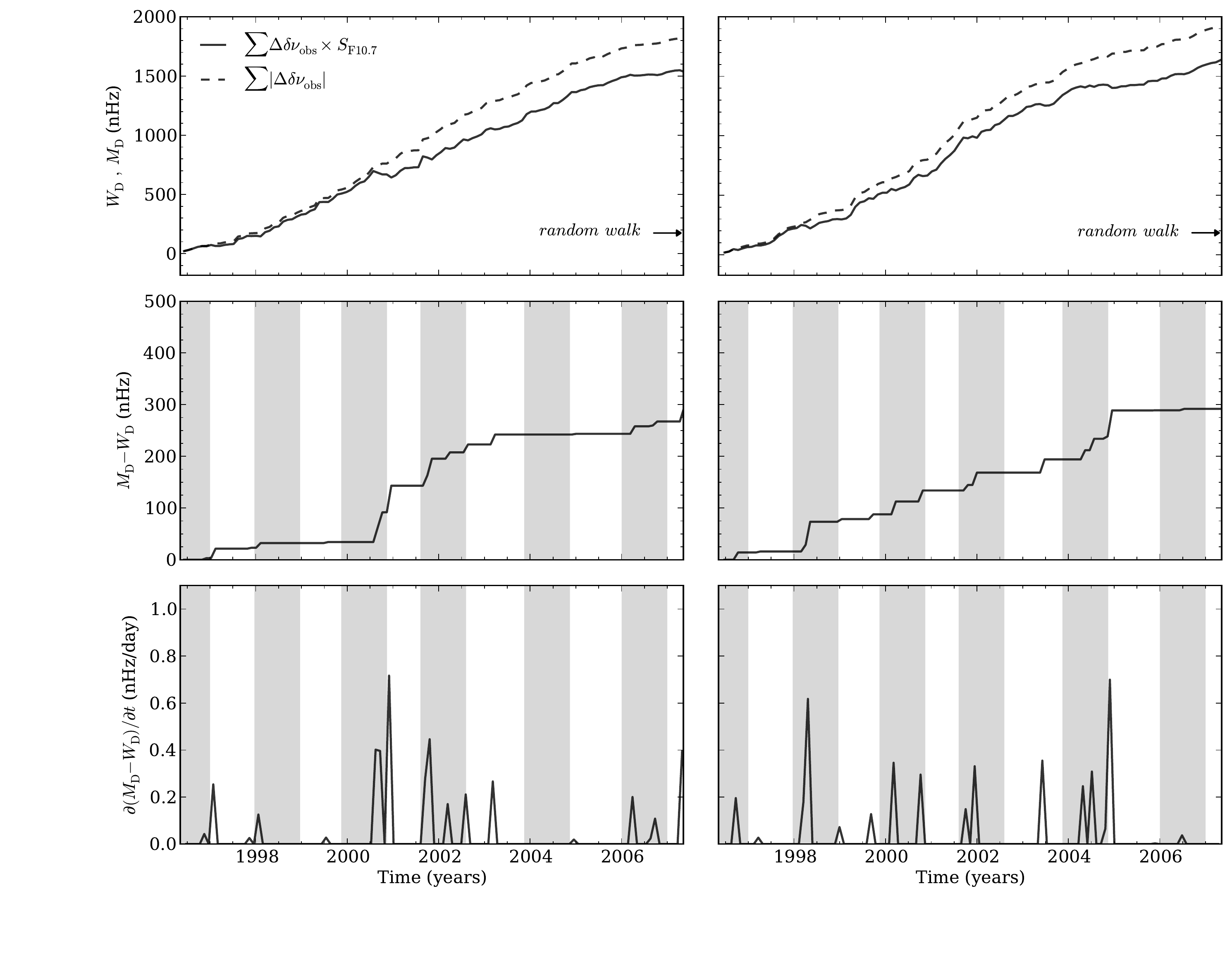}
\caption{The left panel concerns Sample 1, while the right panel corresponds to Sample 2. Top panels: Weighted sum of the variations in the observed frequency shifts (black solid line), where the weight $S$, in equation (\ref{eq:cortest}), is determined by the variation in the 10.7cm flux. The dashed line represents the sum of the absolute (modulus) values of the frequency-shift differences, $M_{\rm D}$. Middle panels: Difference between the dashed and solid lines in the top panels. Bottom panels: Derivative of the difference shown in the middle panels. The grey bars, with a width of one year, are centered around the approximate location of the maxima of the quasi-biennial signal detected by \citet{Broomhall2012}.}\label{fig:flux}
\end{figure*}

The acoustic frequency shifts are found to be better correlated with the 10.7cm flux than with other activity proxies \citep[e.g.][]{Chaplin2007,Tripathy2007,Jain2009}. Besides the contribution from sunspots, the 10.7cm flux is also affected by radio plages and the quiet-Sun background in the upper chromosphere and lower corona \citep[e.g.][]{Covington1969,Tapping1987,Tapping1990}. Given its sensitivity to both sunspots and plages, the 10.7cm flux may provide a more complete picture of the short-term variations in the solar activity than the sunspot areas alone. To test this possibility, we repeated our analysis using the variations in the 10.7cm flux (from NGDC/NOAA) instead of the sunspot area variations as the weight in equation (\ref{eq:cortest}). The results, shown in Fig. \ref{fig:flux}, confirm that the loss of correlation is less pronounced in this case than when the sunspot area variations are used (around 40 per cent smaller than that found in Figs. \ref{fig:resobs} and \ref{fig:diffderiv}). Despite this, we still find that, in general, the loss of correlation is more significant around the maxima of the quasi-biennial variations.

The comparison between the results found for the area covered by sunspots and for the 10.7cm flux confirms that the latter contains contributions from additional activity-related features (besides sunspots) that vary on short time-scales and that these features have a significant impact on the short-term variations of the frequency shifts. Moreover, we may conclude from the results for the 10.7cm that spots and plages alone do not fully explain the short-term variations in the observed frequency shifts.

\section{Conclusions}\label{sec:conclusions}

In this work we investigated the correlation between the variation in the solar frequency shifts and in the area covered by sunspots. In particular, we proposed a new observable, consisting in the sum of the frequency shifts weighted by the variation of the sunspot area, which isolates and amplifies the signal from the short-term variations in the frequency shifts. 
Using this new observable, we found a strong correlation between the short-term variations in the area covered by sunspots and in the frequency shifts. Nevertheless, a significant loss of correlation is still observed, generally coinciding with the times of maxima of the quasi-biennial variations seen in the solar acoustic frequencies.
The loss of correlation on short time-scales suggests that other physical phenomena, besides sunspots, acting on time-scales shorter than 36 d contribute to the frequency shifts and that their relative importance changes in phase with the quasi-biennial signal.

We also considered the case in which the variation of the 10.7cm flux, rather than the visible sunspot area, is used as a weight in the computation of the new observable. The short-term variation in the frequency shifts was found to vary even more closely in line with the 10.7cm flux than with the sunspot areas, confirming that the 10.7cm flux contains information about additional activity-related features which contribute to the frequency shifts. Nevertheless, a significant loss of correlation, whose physical origin remains to be fully understood, is again observed around the times of maxima of the quasi-biennial signal.

\section*{Acknowledgements}

This work was supported by Funda\c{c}\~{a}o para a Ci\^{e}ncia e a Tecnologia (FCT) through the research grant UID/FIS/04434/2013. ARGS acknowledges the support from FCT through the Fellowship SFRH/BD/88032/2012 and from the University of Birmingham. MSC and PPA acknowledge support from FCT through the Investigador FCT Contracts No. IF/00894/2012 and IF/00863/2012 and POPH/FSE (EC) by FEDER funding through the programme Programa Operacional de Factores de Competitividade (COMPETE). TLC and WJC acknowledge the support of the UK Science and Technology Facilities Council (STFC). The research leading to these results has received funding from EC, under FP7, through the grant agreement FP7-SPACE-2012-312844 and PIRSES-GA-2010-269194. ARGS, MSC, and PPA are grateful for the support from the High Altitude Observatory (NCAR/UCAR), where part of the current work was developed.




\bibliographystyle{mnras}
\bibliography{shortv} 

\bsp  
\label{lastpage}
\end{document}